\documentclass[final]{svjour3}
\usepackage{graphicx}
\usepackage{rotating}
\usepackage{amssymb}
\usepackage{mathptmx}
\usepackage[sort, numbers, compress]{natbib}
\makeatletter
\journalname{Journal of Low Temperature Physics}
\bibpunct{[}{]}{,}{n}{}{,}

\begin{document}

\newcommand{\hdblarrow}{H\makebox[0.9ex][l]{$\downdownarrows$}-}
\title{Pushing the Limits of Broadband and High Frequency Metamaterial Silicon Antireflection Coatings}

%\and K.T. Crowley \and K.H. Miller \and S.M.Simon, B.J. Koopman}

\author{K.P. Coughlin \and J.J. McMahon \and K.T. Crowley \and B.J. Koopman \and K.H. Miller \and S.M. Simon \and E.J. Wollack}

\institute{K.P. Coughlin \email{kpcoughl@umich.edu} \and J.J. McMahon \and S.M. Simon \at Department of Physics, University of Michigan, 450 Church St., Ann Arbor, MI, 48109 \and K.T. Crowley \at Department of Physics, Princeton University, Princeton, NJ, 08544 \and B.J. Koopman \at Department of Physics, Cornell University, Ithaca, NY 14850 \and K.H. Miller \and E.J. Wollack \at NASA Goddard Space Flight Center, Greenbelt, MD, 20771}

\maketitle

\begin{abstract}

Broadband refractive optics realized from high index materials provide compelling design solutions for the next generation of observatories for the Cosmic Microwave Background (CMB), and for sub-millimeter astronomy. In this paper, work is presented which extends the state of the art in silicon lenses with metamaterial antireflection (AR) coatings towards larger bandwidth and higher frequency operation. Examples presented include octave bandwidth coatings with less than $0.5\%$ reflection, a prototype 4:1 bandwidth coating, and a coating optimized for 1.4 THz. For these coatings the detailed design, fabrication and testing processes are described as well as the inherent performance trade offs.


\end{abstract}

\section{Introduction}

Current and planned ground-based CMB experiments  \cite{ADVACTPol,PB,SPT3G,CLASS,BICEP3} and sub-millimeter observatories \cite{SCUBA,SHARC} use photon noise limited detectors.  Improving the sensitivity of these instruments requires deploying a greater number of detectors.  The two approaches being pursued are development of optical systems with large fields of view, and multichroic detectors.  The former usually requires lenses fabricated from high index of refraction materials.  The latter requires optical systems operating over broad bandwidths.  Simultaneously meeting these design requirements necessitates the realization of high performance AR coatings.

Silicon lenses with metamaterial AR coatings represent a rapidly maturing technology that can meet these needs.  These coatings offer excellent optical performance (low reflection, low dielectric losses, and low scattering \cite{Datta:13}), are robust to thermal cycling, and can be reliably manufactured.  Moreover, these coatings offer fine control over dielectric properties. By tailoring the subwavelength features in the surface of the silicon, the dielectric constant can be tuned to a broad range of values between pure silicon and vacuum \cite{Datta:13}.  This freedom allows these coatings to be optimized for a wide variety of applications.

In this paper, we present work done to explore the range of applications possible with silicon dielectric metamaterial AR coatings.  The paper is organized as follows: Section 2 presents the design and fabrication methods for these coatings.  Section 3 discusses testing and performance.  Section 4 presents three types of coatings: 3 layer with 3:1 bandwidth from 75-170\,\,GHz and 125-280\,\,GHz ; 5 layer with 4:1 bandwidth from 75-320\,\,GHz; and a single layer coating centered at 1.4 THz.  Section 5 concludes with discussion of the applicability of this coating technology, including the trade-offs between broad bandwidth and high frequency operation.

\section{Design and Fabrication}

The metamaterial AR coatings discussed here are periodic structures fabricated with silicon dicing blades.  Figure \ref{fig:hfss_model_all} shows the dielectric metamaterial unit cells corresponding to 1, 3 and 5 layer AR coatings, as modeled in Ansys High Frequency Structural Simulator (HFSS). These structures are fabricated by making arrays of evenly spaced grooves in the silicon surface, rotating the substrate by 90 degrees, and repeating.  By changing the blade thickness and spacing of the cuts, one can tune the dielectric constant of each metamaterial layer.  Creating multilayer coatings requires realizing nested cuts with multiple widths.  These coatings perform exactly like layers of dielectric up to a high frequency limit that is set by the pitch of the unit cell \cite{Rytov:56,Egan:82,Raguin:93}.

The design of these AR coatings requires tuning of a number of free parameters and adhering to practical constraints.  The free parameters are the pitch of the cuts, kerf (kerf is the term that describes the width of the cut) and depth of each layer, giving $2n+1$ free parameters for an n-layer coating.  The constraints come from particulars of the fabrication system approach and scattering as discussed in the next section. 

\subsection{Constraints}

The primary constraint comes from the blades which are manufactured with a maximum exposure-to-kerf ratio of 50:1.  This limits the ratio of the depth to the widths of the cuts.  For example, 20 $\mu$m wide cuts must be below 1\,\,mm in depth.  Beyond this limit, the blade wanders while cutting, which increases pillar breakage, wear, and probability of breakage of the blade, and can increase scattering.  This constraint limits how many layers the AR coating can have and limits the maximum effective index of the deepest AR layer.    

A second effect limits the kerfs of successive layers.  While the cut profiles are usually quite sharp and square, there is always some taper to the cuts.  If two successive cuts are too close in intended kerf, the interface between the two cuts begin to blend together, reducing performance.  
There are also practical limitations to the minimum achievable blade thickness. Presently, the diamond-nickel blades we use are available down to 12 $\mu$m in thickness. This limits the frequency of single layer coatings to approximately 2 THz.

A fundamental constraint of these metamaterials is that the defined features need to be sufficiently small that they act as a simple dielectric material.  This requires the pitch of the cuts to be less than about half the wavelength \cite{Egan:82, Raguin:93}.  For a given geometry, increasing the frequency beyond this breakdown frequency will lead to scattering and a rapid degradation in performance.

\subsection{Optimization}

The design process is a three step process following Datta et al \cite{Datta:13}. The first step uses an analytic transmission line model to perform a Monte Carlo minimization of in-band reflection. This model treats each metamaterial layer as an ideal dielectric.  Next, an empirical model is used to convert the tuned indicies of refraction of each layer into a fill fraction to follow the effective mean field theory from Rytov \cite{Rytov:56}.  While this works relatively well for single layers, its performance degrades for multilayer coatings \cite{Datta:13}.  Despite this, it serves as a useful starting point for the final numerical optimization.  

The last step in the optimization uses numerical modeling in HFSS.  The calculation represents the coating as a single unit cell with periodic boundaries and floquet ports which simulate an infinite array.  The prior step design parameters are used as initial conditions.  HFSS's optimization algorithms are used to minimize the reflection across the band, subject to the fabrication constraints described above.

We used this procedure to design three coatings which are presented here: (1) a 3 layer octave bandwidth coatings prototype of 75-170 GHz and 125-280 GHz bands (2) a 5 layer 4:1 ratio bandwidth coating optimized using blades we had on hand, and (3) a single layer coating for 1.4 THz.  

\begin{figure}
\begin{center}
\includegraphics[height=2in, keepaspectratio]{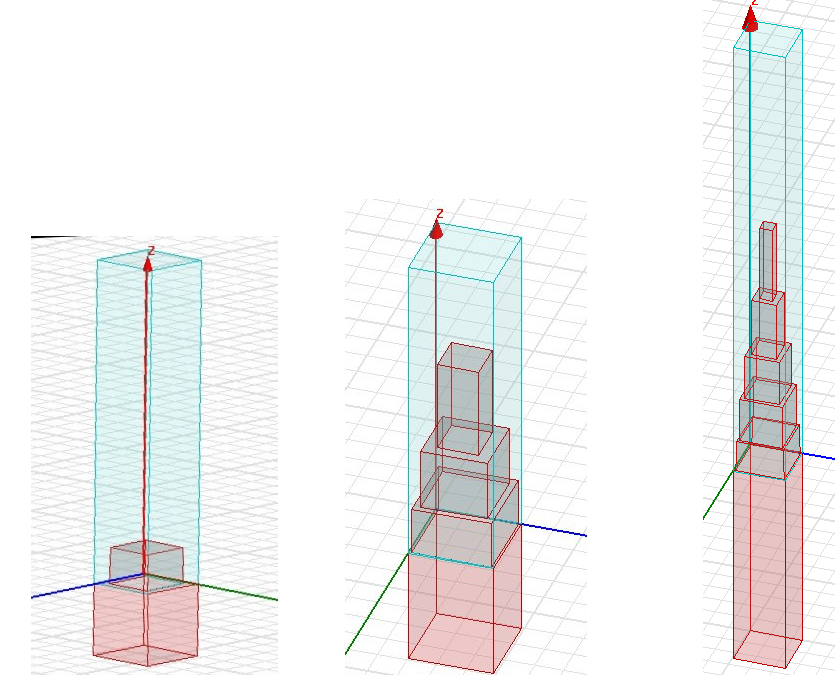}
\caption{HFSS models of the 1, 3, and 5 layer coatings.}
\label{fig:hfss_model_all}
\end{center}
\end{figure}

For the high-frequency coating, the kerf was fixed at 12 $\mu$m, the minimum blade size, allowing only the pitch and depth to vary.  For the 5 layer, we constrain each of the cut widths to the kerf of blades that we currently have in stock.  The minimum blade width for the deepest cut was fixed at 20 $\mu$m. The total depth was kept below the minimum of 1 mm.  The final designs are shown in Figure \ref{fig:hfss_model_all}.

\begin{footnotesize}
\begin{table}
    \begin{center}
    \begin{tabular}{| l | l | l | l |}
    \hline
    Frequency Range & Pitch ($\mu$m)& Kerf ($\mu$m)& Depth ($\mu$m)\\ \hline
    1 Layer 125-165 GHz & 200 & 80 & 250 \\ \hline
    1 Layer 1.3-1.5 THz & 30 & 12 & 25 \\ \hline
    3 Layer 75-170 GHz & 450 & 230 & 500 \\
     & & 110 & 320 \\
     & & 25 & 252 \\ \hline
    3 Layer 125-280 GHz & 285 & 168 & 296\\
    & & 82 & 187\\
    & & 22 & 143\\ \hline
    5 Layer 70-350 GHz & 225 & 152 & 280 \\
     & & 129 & 270 \\
     & & 80 & 165\\
     & & 42 & 150\\
     & & 20 & 125\\

    \hline
    \end{tabular}
    \end{center}
    \caption{This table lists the cut parameters for the various AR coatings which have been fabricated using our dicing saw}
    \label{param_table}
\end{table}
\end{footnotesize}

\subsection{Fabrication}

These designs are fabricated using a custom 3-axis dicing saw \cite{Munson:Thesis}.  A $\mu$m precision contact metrology probe is used to measure the surface of the silicon.  A surface model is then constructed and used to create cut paths which are followed by the saw.

The cutting is done using diamond-embedded nickel dicing blades of varying thicknesses.  For a single layer AR coating, only a single blade is used.
Before and after each set of cuts is made, we made a test cut in a separate silicon wafer and examined the blade profile to ensure that the kerf is within tolerance.  By tracking the behavior of the blade as it cuts, we can better understand our fabrication process, and more accurately predict the overall performance of the coating.  For most blades, the kerf changes by less than 2 $\mu$m while cutting a 30\,\,cm lens ($\sim 100$ m of cuts).  

For a multilayer geometry, the cutting is typically done thickest to thinnest blade.  This is preferred so that as each blade cuts, the wear across the blade is uniform.  For the two thinnest cuts for the 5 layer (40 $\mu$m and 20 $\mu$m widths), the cut order is reversed.  The profile of the thinnest blade degrades significantly near the top of the cut (to about 50 $\mu$m depth).  By reversing the order of the cuts, this degraded part of the cut is removed by the 40 $\mu$m wide layer.

For our 75-170 GHz 3-layer design, we had to again change our cutting technique.  The cut for the top layer of the coating is 230 $\mu$m wide, 500 $\mu$m deep.  Using this thickness blade for this cut proved to be too much of a strain on our gantry.  Our solution was to use multiple passes to make the cut.  We made two cuts using an 80 $\mu$m wide blade to define the edges of the cut, and then came in with a 180 $\mu$m wide blade to clear out the remaining silicon.  This results in a very sharply defined layer without causing undue stress on our gantry.

The overall yield of the fabrication is very high, with fewer than one pillar broken per 100,000 on the 5 layer coating, and zero broken pillars on the high frequency coating.  The largest lenses done to date are just over 30 cm, but our fabrication system can handle up to 46 cm lenses which is set by the size of available single-crystal silicon.  We don't foresee any new issues going to larger size silicon.

\section{Testing}

These prototypes were tested using a combination of a coherent reflectometer below 300 GHz and a Fourier Transform Spectrometer (FTS) for the higher frequencies.

\begin{figure}
	\begin{center}
	\includegraphics[width=0.8\textwidth,keepaspectratio]{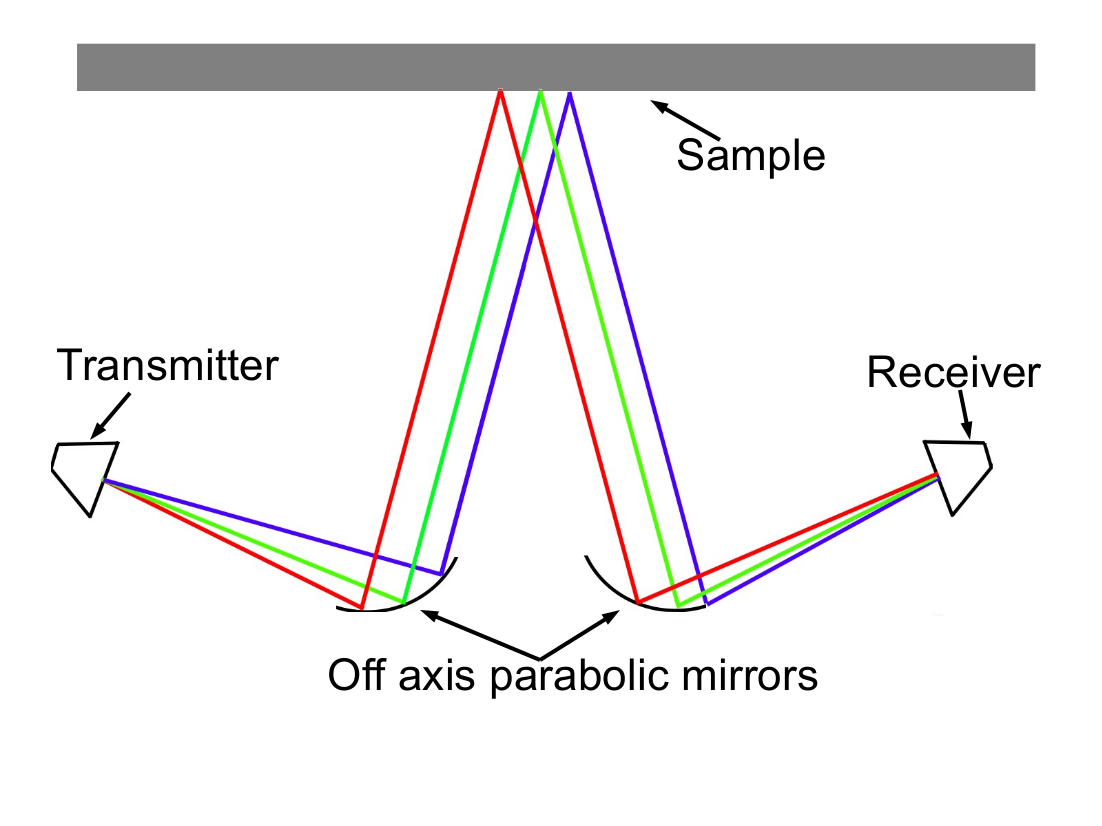}

	\caption{A schematic diagram of our reflection measurement set up.  The first collimating mirror directs the beam at a 10 degree angle of incidence.  The second mirror collects the reflected beam, and has an absorbing stop above it, limiting the aperture before directing the beam to the detector.}
	\label{fig:reflectometer}
	\end{center}
	
\end{figure}

The reflectometer setup is depicted in Figure \ref{fig:reflectometer}. The core of the system is a Virginia Diodes active multiplier chain operates from 75-250 GHz.  A feed horn couples this source to a collimating mirror, which directs the beam to the sample to be measured.  The sample is mounted horizontally on a three point mount.  Each point can be adjusted using a precision micrometer.  An aperture-limiting stop is placed above an identical mirror, which then collects the reflected light and directs it to an identical feed horn, coupled to a detector diode.  This detector feed is mounted on a three axis mount, which is used to adjust the position of the feed to maximize the signal.  The diode is readout using a lock-in amplifier with the source chopped at 200 Hz.  By comparing the AR coated surface to a reflective surface, we can get a measurement for the fractional power reflected.  The beam size of our measurement is 3 inches.  We test the uniformity of the coating by measuring various spots on the surface of the plate.

\section{Results and Discussion}

In this section, we discuss the measured performance of these AR coating prototypes.

\subsection{3 layer coating}

\begin{figure}[htbp]
\begin{center}
\includegraphics[width=0.8\linewidth, keepaspectratio]{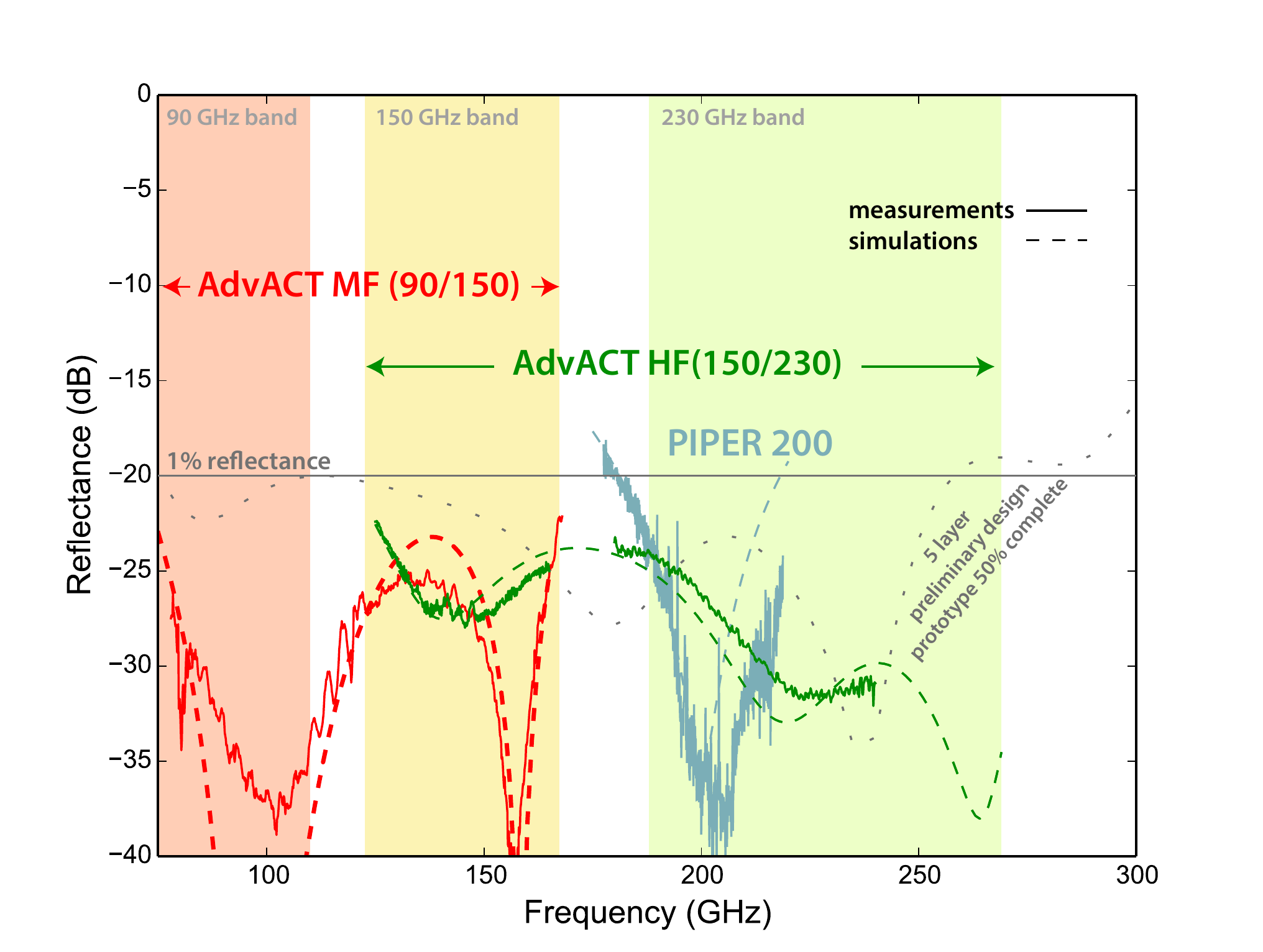}

\caption{Plot of reflection as a function of frequency, showing several previously made AR coatings.  The coatings all show good performance, as well as excellent agreement with predicted theory.}
\label{fig:refl_all}
\end{center}

\end{figure}

We fabricated 3 layer AR coatings covering the 75-170 GHz and 125-280 GHz bands for the Advanced ACTPol project \cite{ADVACTPol}.  The reflection data for these is shown in Figure \ref{fig:refl_all}.  Both of those designs achieved octave bandwidth with subpercent reflection and excellent agreement with predicted reflection.  Based on a loss tangent of 7e-4 (our upper limit for loss tangent of this material) we predict a dielectric loss of less than $0.5\%$ for each coating.  These coatings represent a mature, high performance broad bandwidth application of this technology.

\subsection{5 layer coating}
\begin{figure}
\begin{center}
\includegraphics[width=1.\textwidth,keepaspectratio]{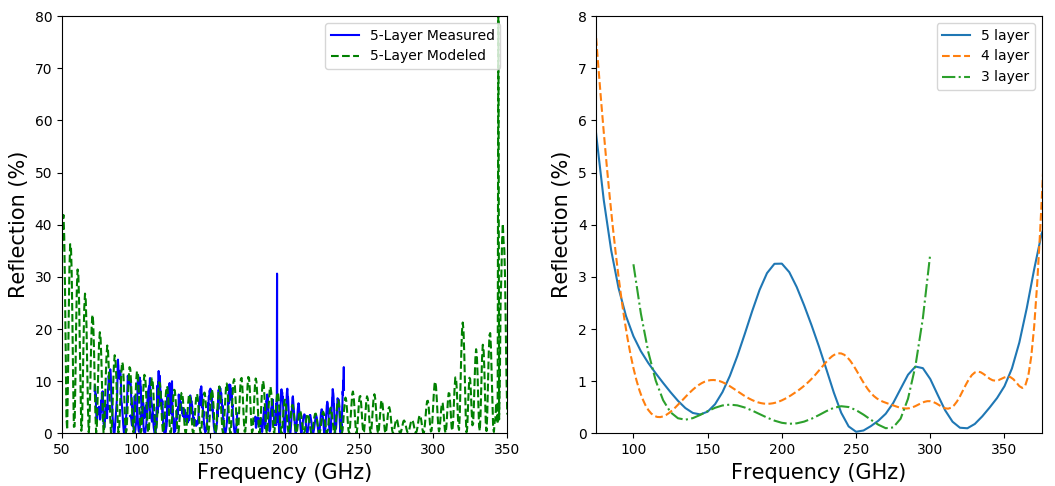}

\caption{(Left) Comparison between the modeled and measured reflection performance for the two sided 5-layer AR coating prototype.  While the reflections are quite large at some frequencies, this serves as a proof of principle, that the computer simulation of this structure matches very well with the real fabricated version.  (Right) Comparison of 4- and 5-layer AR coatings.  These are computer simulations of single sided AR structures constrained by our fabrication system.  As can be seen, the 5-layer does not significantly outperform the 4-layer.  This is mainly due to the depth constraint.  By having a hard limit on the depth of the thinnest cut (at 50x the width of that cut), each layer is made slightly thinner, degrading the performance.  With one fewer layer, the 4-layer allows for better optimization of the depth of each layer, recovering performance lost by losing a layer.}
\label{fig:five_layer}
\end{center}
\end{figure}

The 5-layer coating was fabricated on both sides of a 3 cm thick piece of low loss silicon.  The results from the reflection measurement are shown in Figure \ref{fig:five_layer}.  The figure shows that in the frequency ranges the coating was measured, it follows the modeled behavior very closely.  

This prototype coating shows that these coatings can be produced in relevant sizes (22 cm diameter) and perform as modeled.  Further, optimization of these blades widths was subject to what was available in lab.  Further optimization could lead to significantly improved performance.

From our experience making this 5 layer AR coating, we find it is impractical to go to any higher number of layers.  Since the dicing blades have a limit on the kerf-to-exposure ratio, we cannot fully optimize our 5 layer design, in that each layer would ideally be deeper than our constraints allow.  Going down to a four layer allows for each layer to be thicker, and better optimized, thus performance does not significantly decrease with one less layer.  Additionally, taking away a layer significantly eases the fabrication.  It takes less time to make, as well as reducing the chance of two layers blending together, as with fewer layers the kerfs are generally more distinct.  For the four-layer design shown in Figure \ref{fig:five_layer}, the final two layers are at 80 $\mu$m and 20 $\mu$m, so the difference between these two kerfs it three times larger than the difference between the last two cuts of the 5 layer.  We therefore argue that 4:1 bandwidth realized with a 4 layer coating is near the limit for this technology.

\subsection{Single Layer Coating}

Single layer AR coatings have been fabricated both on lenses (concave-convex) and on flat prototypes.  Having only a single layer effectively fixes the bandwidth of the coating.  We have previously fabricated a single layer AR coating for the PIPER balloon CMB experiment.  It was a narrow band coating centered at 200 GHz.  The reflection for this coating was measured with our reflectometer and is shown in Figure \ref{fig:refl_all}

\begin{figure}
\begin{center}
\includegraphics[width=0.8\textwidth,keepaspectratio]{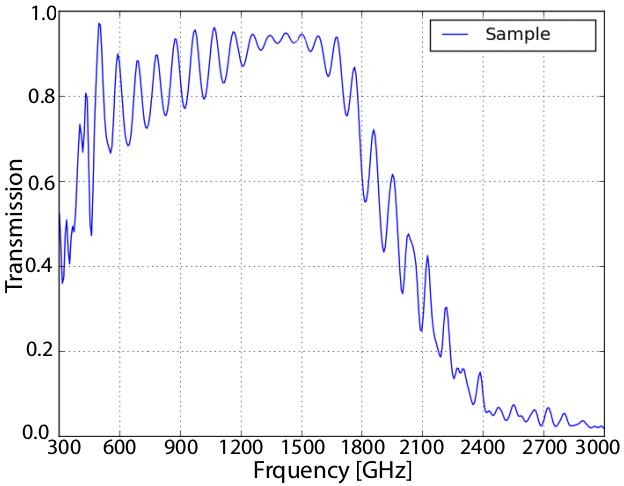}
\caption{Plot of transmission of our one inch high-frequency AR coating as measured by an FTS.  This shows the transmission of the coating peaks at 1.43 THz.  The FTS used a 125 $\mu$m mylar beamspliter and the source was a mercury arc lamp.}
\label{fig:THz_mesurement}
\end{center}

\end{figure}

The high frequency single layer AR coating was optimized at 1.4 THz.  The transmission performance of a coated 1 inch silicon wafer as measured by an FTS is show in Figure \ref{fig:THz_mesurement}. It is possible to go to slightly higher frequencies.  By decreasing the pitch, we can increase the breakdown frequency.  However, since decreasing the pitch changes the fill fraction, and we cannot compensate by making the blade any smaller, this limits what the effective index of the simulated dielectric can be, and thus the performance of the AR coating begins to degrade past about 2 THz.  That said, the performance is still reasonable ($\sim 2\%$ reflection) as high as 3 THz.

Additionally, there is a limitation to the precision of our dicing saw.  We can control the depth of our cuts to within less than 5 $\mu$m, but at such high frequencies, even this small error can degrade the performance.  

\section{Conclusion}

We have explored the frequency and bandwidth scalability of this technology.  We have presented results from a fully optimized octave bandwidth 3 layer coating, and proof of principle measurements of an extreme bandwidth 5 layer coating and a high frequency 1.4 THz coating. 

This work demonstrates that metamaterial silicon AR coatings can realize low reflectance with broad bandwidths and high frequency.  These demonstrated test cases span the space of applicability of these coatings fabricated with silicon dicing saws.  The bandwidth is limited by our ability to produce high aspect ratio multilayer structures, the availability of blades narrow enough for a given high frequency limit for the band, and the criteria for optimization.  Thus the exact limits live in a complex space.  Practical constraints conspire to favor four layer coatings as the maximum number worth implementing.  If reflectance of $2\%$ can be tolerated, we find that we can achieve 4:1 bandwidth with 4 layers operating up to 350 GHz.  Extending to higher frequencies requires narrower blades with large exposures.  This limits the number of layers possible as the frequency increases. With fewer layers, high frequency is possible as we showed with the 1.4 THz single layer prototype.  Provided one navigates the optimization and constraints procedure with great care, metamaterial silicon offers excellent performance over a wide range of frequencies and bandwidths.  

\begin{acknowledgements}
This work was supported by a NASA Office of the Chief Technologist’s Space Technology Research Fellowship \# NNX15AP46H. McMahon Lab efforts were supported by 
NNX13AE56G, NNX14AB58G, and DE-SC0015799
\end{acknowledgements}

\pagebreak

\end{document}